\documentclass[aps,twocolumn,showpacs,amsmath,amssymb]{revtex4}
\usepackage{dcolumn}
\usepackage{bm}
\usepackage{color}
\usepackage{latexsym}
\usepackage{amssymb}
\usepackage{graphicx}
\usepackage{ulem}

\begin{document}
\title{Circular Hall Effect in a wire}
\author{M. V. Cheremisin}
\affiliation{A.F.Ioffe Physical-Technical Institute,
St.Petersburg, Russia}
\date{\today}
\begin{abstract}
A constant longitudinal current in a wire is accompanied by azimuthal magnetic field.
In absence of the radial current in a wire bulk the nonzero radial(Hall) electric field must be present. The
longitudinal current can be viewed as collective drift of carriers in crossed magnetic(azimuthal) and electric(radial) fields,
hence can be ascribed as Circular Hall Effect. At low temperatures the enhanced carrier viscosity leads to nonuniform current
density whose radial profile is sensitive to presence of $collinear$ diamagnetic currents nearby the wire inner wall.
Both the current and azimuthal magnetic field are squeezed out from the bulk towards the inner wall of a wire. Magnetic
properties of a sample resembles those expected for ideal diamagnet. At certain critical temperature a former dissipative
current becomes purely diamagnetic providing the zero resistance state. At low currents the temperature threshold is found
for arbitrary disorder strength and the sample size. For bulky sample and finite currents the threshold temperature is
found as a function of the magnetic field.
\end{abstract}
\maketitle

\section{Introduction}
\label{Introduction}
Usually, the Hall measurements\cite{Hall1879} imply the presence of external magnetic field source.
Evidence shows that the current itself produces a finite magnetic field which may, in turn, influence the current carrying state.
In the present paper, we take the interest in a special case when the only current-induced magnetic field is present. Therefore, we
reveal a Circle Hall Effect in a round cross section conductor. The account of finite carrier viscosity and diamagnetic
currents at the inner boundary of the wire provides a certain feasibility of the zero resistance state at low temperatures.

\section{Circular Hall Effect: uniform current flow}
\label{Circular Hall Effect: uniform current flow}

The conventional Drude equation for 3D electrons placed in arbitrary electric $\mathbf{E}$ and magnetic $\mathbf{B}$ fields yields
\begin{equation}
\frac{\partial \mathbf{V}}{\partial t}=\frac{e \mathbf{E}}{m}+\left [\mathbf{V}\times \mathbf{\Omega}_{c} \right]-\frac{\mathbf{V}}{\tau},
\label{Drude}\\
\end{equation}
where $e$ is absolute value of the electric charge, $\mathbf{\Omega}_{c}=\frac{e\mathbf{B}}{mc}$ is the cyclotron frequency vector, $m$ is the effective mass, $\tau$ is the momentum relaxation time due to collisions with impurities and(or) phonons, $\mathbf{V}$ is the carrier flux velocity.

For steady state one obtains the following equation
\begin{equation}
\mathbf{V}=\mu \mathbf{E}+\left [\mathbf{V} \times \mathbf{\Omega}_{c}\tau \right],
\label{vector equation}\\
\end{equation}
where $\mu=\frac{e\tau}{m}$ is the carrier mobility. For arbitrary orientation of the electric
and the magnetic fields the exact solution of Eq.(\ref{vector equation}) is straightforward \cite{Anselm78}.

\begin{figure}[tbp]
\begin{center}\leavevmode
\includegraphics[width=0.7\linewidth]{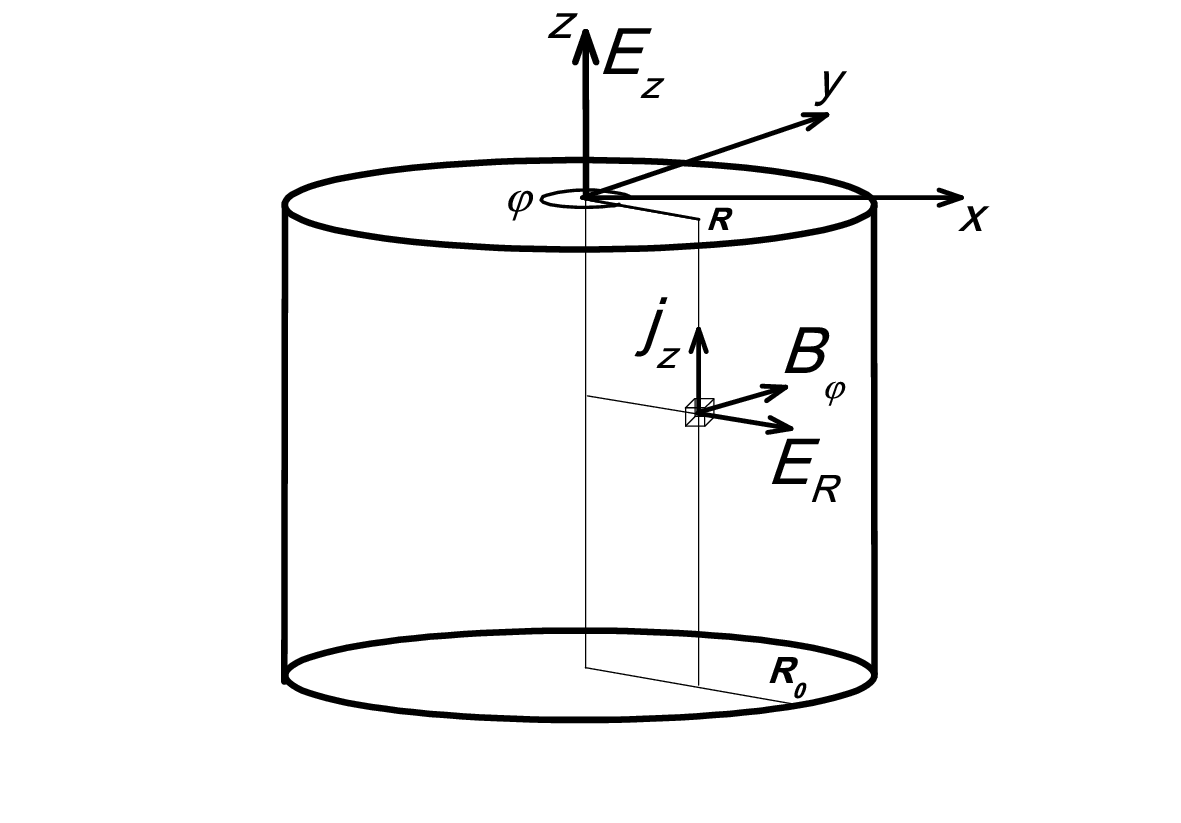} \caption[]{\label{Fig1} Schematic view of the Circular Hall effect.}
\end{center}
\end{figure}

Let us restrict ourself to a wire of radius $R_{0}$ and, hence use the cylindrical
geometry frame. The voltage source(not shown in Fig.\ref{Fig1}) is attached to a free wire ends providing the
longitudinal electric field $E_{z}$. The carrier velocity $v_{z}$ is uniform. Following to Biot-Savart law the longitudinal
current density $j_{z}=neV_{z}$ results in azimuthal magnetic field $B_{\varphi}=\frac{2\pi j_{z}}{c}R$, where $R$ is
the radial coordinate, $n$ is the carried density. The azimuthal magnetic
field reaches the maximal value $B_{0}=B_{\varphi}(R_{0})$ at the rod wall. Note that the radial current is absent
in the sample bulk. Hence, the nonzero radial electric field $E_{R}$ must exist to prevent Lorentz force
$\sim V_{z}B_{\varphi}/c$ action. Evidently, the radial electric field plays the role of Hall
one regarding conventional description\cite{Hall1879}.

Following the above reasoning we re-write Eq.(\ref{vector equation}) for both the longitudinal $v_{z}$ and radial $v_{r}=0$ components
of the carrier velocity as it follows:
\begin{eqnarray}
V_{z}=\mu E_{z},
\label{velocity_Drude}\\
V_{z}=c\frac{E_{R}}{B_{\varphi}}.
\label{velocity_Drude+}
\end{eqnarray}
Eq.(\ref{velocity_Drude}) provides a familiar differential Ohm's law. By contrast, Eq.(\ref{velocity_Drude+})
presents the novel view on the longitudinal current as a carriers drift in crossed $E_{R}\perp B_{\varphi}$ fields, i.e.
ascribes a Circular Hall Effect. We argue that the radial electric field $E_{R}=2\pi neR\left(\frac{V_{z}}{c}\right)^{2}$
defines volumetric charge density $Q=\text{div}\mathbf{E}/4\pi=ne\left(\frac{V_{z}}{c}\right)^{2}$. Thus, the wire is charged\cite{Matzek68,McDonald2010}
since $Q/ne\ll 1$.

\section{Circular Hall Effect: Nonuniform viscose flow}
\label{Nonuniform current flow}
We now intend to answer a question whether the current carrying state in a wire can be nonuniform in
radial direction, namely $V_{z}(R)$. Navier-Stokes equation modified with respect to presence of the magnetic field yields
\begin{equation}
\frac{\partial \mathbf{V}}{\partial t}+(\mathbf{V}\nabla)\mathbf{V}=\frac{e \mathbf{E}}{m}+\left [\mathbf{V}\times \mathbf{\Omega}_{c} \right]+\hat{\eta} \triangle \mathbf{V}-\frac{\mathbf{V}}{\tau}.
\label{Navier-Stokes}\
\end{equation}
Here, $\hat{\eta}$ is the viscosity tensor\cite{Steinberg58,Alekseev16} whose longitudinal and transverse components

\begin{eqnarray}
\eta_{xx}=\eta_{yy}=\frac{\eta}{1+4\Omega_{c}^{2}\tau_{ee}^{2}},
\label{viscosity_components}\\
\eta_{xy}=-\eta_{yx}=\eta_{xx}2\Omega_{c}\tau_{ee}
\nonumber
\end{eqnarray}
depend on magnetic field. Then, $\eta=\frac{1}{5} v^{2}_{F}\tau_{ee}$ is
the kinematic viscosity of the carriers at zero magnetic field, $V_{F}$ is the Fermi velocity, $\tau_{ee}$ is the electron-electron collisions time. Viscosity effects become important\cite{Gurzhi63} when the e-e mean free path $l_{ee}=V_{F}\tau_{ee}$ is less and(or) comparable to that caused by phonons and(or) impurities $l=V_{F}\tau$ and typical length scale of the sample. Note that under assumption of radially dependent velocity $V_{z}(R)$ the Euler
term can be neglected in Eq.(\ref{Navier-Stokes}).

For steady state Eq.(\ref{Navier-Stokes}) can be re-written for both the longitudinal and radial components of carrier velocity as it follows
\begin{eqnarray}
\eta_{xx} \tau \triangle_{R} V_{z}-V_{z}+\mu E_{z}=0,
\label{XX_velocity_Stokes}\\
\eta_{yx} \triangle_{R}V_{z}+\frac{eE_{R}}{m}-\frac{V_{z}eB_{\varphi}}{mc}=0.
\label{YX_velocity_Stokes}\\
\nonumber
\end{eqnarray}
Here, $\triangle_{R}=\frac{1}{R}\frac{d}{dR}\left (R \frac{d}{dR}\right )$ is the radial component of the Laplace operator.
Our primary interest concerns Eq.(\ref{XX_velocity_Stokes}) which determines the nonuniform velocity profile $V_{z}(R)$
and, in turn, the azimuthal magnetic field
\begin{equation}
B_{\varphi}=\frac{4\pi ne}{c R}\int\limits^{R}_{0}V_{z}(R)R dR.
\label{Azimuthal_Field}\
\end{equation}
Introducing the dimensionless velocity $v=V_{z}(R)/\mu E_{z}$ and the reduced radius $r=R/R_{0}$, one may rewrite
Eq.(\ref{XX_velocity_Stokes}) as it follows
\begin{equation}
\frac{\eta_{xx}}{\eta} \nu^{-2} \triangle_{r}v-v+1=0.
\label{Vz_Stokes_Modified}\\
\end{equation}
where $\nu=R_{0}/\lambda$ is the dimensionless parameter, $\lambda=\sqrt{\eta \tau}$ is the typical length scale of the problem.
The condition $\nu \ll 1$( $\nu \gg 1$ ) determines the high(low)-viscous electron gas respectively.

We argue the solution of Eq.(\ref{Vz_Stokes_Modified}) is complicated due to magnetic field dependent longitudinal viscosity.
In principle, Eq.(\ref{Vz_Stokes_Modified}) can be expressed in terms of the reduced magnetic field $B_{\varphi}/B_{0}$ via
relationship $v=\frac{1}{2r}\frac{d}{dr}(r B_{\varphi}/B_{0})$ but still remains difficult for analytical processing. Fortunately,
at small currents and(or) small magnetic fields $\Omega_{c}\tau_{ee}\ll 1$ the longitudinal viscosity can be kept constant $\eta_{xx}\sim \eta$, thus
Eq.(\ref{Vz_Stokes_Modified}) becomes amenable for analytic analysis. Noteceably, at low magnetic fields the transverse
viscosity can be disregarded in Eq.(\ref{YX_velocity_Stokes}) which, in turn, gives a familiar result for
carrier drift in crossed $E_{R}\perp B_{\varphi}$ fields.

Under the above assumptions the solution of Eq.(\ref{XX_velocity_Stokes}) is straightforward:
\begin{equation}
v(r)=1+C_{1}J_{0}(\nu r)+C_{2}Y_{0}(\nu r),
\label{Solution_Bessel1}\\
\end{equation}
where $J_{0}$ and $Y_{0}$ are zero-order modified Bessel functions of the first and second kind respectively. Since the carrier velocity remains
always finite, we conclude that $C_{2}=0$, because $Y_{0} \rightarrow \infty$ at $r \rightarrow 0$. Introducing a general condition
$v|_{R=R_{0}}=v_{0}$ for longitudinal velocity at the inner wire wall Eq.(\ref{Solution_Bessel1}) yields
\begin{equation}
v(r)=1+(v_{0}-1)\frac{J_{0}\left (\nu r \right )}{J_{0}(\nu)},
\label{Solution_Bessel2}\\
\end{equation}
Note that the trivial case of the uniform current flow examined in Sec.\ref{Circular Hall Effect: uniform current flow}
follows from Eq.(\ref{Solution_Bessel2}) when $v_{0}=1$. We now demonstrate that the boundary condition at the inner wire
surface alters crucially the radial velocity profile and, moreover, the sample resistivity.

\subsection{Poiseuille viscose flow}
\label{Poiseuille viscose flow}
\begin{figure}[tbp]
\begin{center}\leavevmode
\includegraphics[width=0.9\linewidth]{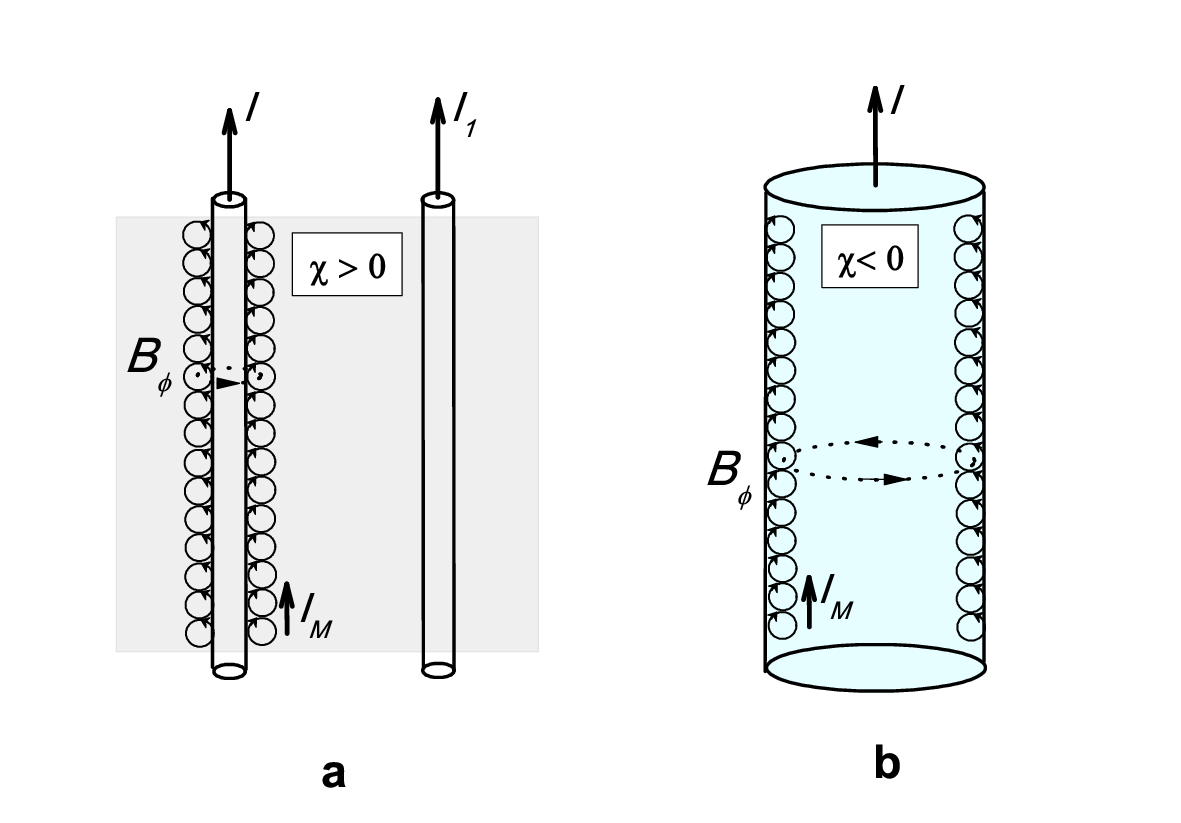} \caption[]{\label{Fig2} The microscopic magnetic currents for a) current carrying conductors placed into paramagnetic $\chi>0$ media ( from Ref.\cite{Sivukhin96}) b) The sole current carrying diamagnetic $\chi<0$ conductor.}
\end{center}
\end{figure}

We recall that the simple wall adhesion condition $v_{0}=0$ \cite{Gurzhi63} could be familiar
regarding the Poiseuille's viscous flow in conventional hydrodynamics. In Fig.\ref{Fig2} the blue curves depict the radial dependence
of the flux velocity $v(r)$ specified by Eq.(\ref{Solution_Bessel2}) for different viscosity strengths. As expected, for small viscosity $\nu \gg 1$ the 3D fluid
velocity is mostly uniform with exception of ultra-narrow layer $\sim \lambda$ close to wire inner wall. In contrast, for highly viscous case $\nu \leq 1$ the flux velocity follows the Poiseuille law
$v(r)=\frac{\nu^{2}}{4}(1-r^{2})$ shown by the dashed line in Fig.\ref{Fig2}.

\subsection{Diamagnetic viscose flow}
\label{Diamagnetic viscose flow}

The special interest of the present paper concerns the possibility of different boundary condition $v_{0}>1$
whose physical background we intend to illustrate hereafter.

At first, recall a scenario of a current carrying wire surrounded by
paramagnetic media(see the left-hand sketch in Fig.\ref{Fig2},a). Let the longitudinal current $I$ is provided by an external source.
The current carrying wire induces the azimuthal magnetic field $B_{\varphi}=\frac{2I}{cR}$ in the surrounding space $R>R_{0}$. Notably,
the magnetic field at the outer wire wall $B_{0}=B_{\varphi}(R_{0})$ results in nonzero microscopic magnetic current $I_{M}=4\pi \chi I$\cite{Sivukhin96,Vlasov05} because of the paramagnetic surrounding. Here, $\chi>0$ is the paramagnetic susceptibility. The total current flowing along
the wire yields $I+I_{M}$. Let an another conductor with a driven current $I_{1}$( see Fig. \ref{Fig2},a ) is placed in parallel to initial one.
Again, the total current along the second wire $(1+4\pi \chi) I_{1}$ includes a microscopic component(not shown in Fig. \ref{Fig2},a) as well.
One can check that Ampere's attractive force $\sim (1+4\pi \chi)I \cdot I_{1}$ between a pair of wires with parallel currents is enhanced by a factor of $1+4\pi\chi$\cite{Sivukhin96} compared to that in absence of paramagnetic media. We conclude that Ampere's force enhancement is caused by
microscopic magnetic currents at the outer wire surface.

We now provide a strong evidence of similar effect for current carrying diamagnetic wire $\chi<0$, shown in Fig.\ref{Fig2},b. Indeed, for certain value
of the applied current $I$ the azimuthal magnetic field at the inner rod surface $B_{0}$ results in extra diamagnetic current $I_{M}$
which is parallel to native current, namely $I_{M}=4\pi|\chi| I$. Phenomenologically, we assume that diamagnetic current may flow
within narrow layer of the width $\delta$. The respective density of diamagnetic current $j_{M}=\frac{I_{M}}{2\pi R_{0} \delta}$ may exceed the ohmic
current density $j_{z}$. One can deduce the dimensionless flux velocity $v_{0}$ at the inner rod surface as
\begin{equation}
v_{0}=\frac{j_{M}}{j_{z}}=\frac{j}{j_{z}}\kappa,
\label{boundary_condition}\\
\end{equation}
where we introduced the average current density $j=\frac{I}{\pi R_{0}^{2}}$ while $j_{z}=ne\mu E_{z}$ is the ohmic current density.
Then, $\kappa=\frac{2\pi R_{0} |\chi|}{\delta}$ is the dimensionless diamagnetic parameter dependent on the sample size. Without
diamagnetic currents, i.e. when $\kappa=0$ we recover the conventional Poiseille's flow
provided by the wall-adhesion condition $v_{0}=0$.

\begin{figure}[tbp]
\begin{center}\leavevmode
\includegraphics[width=0.9\linewidth]{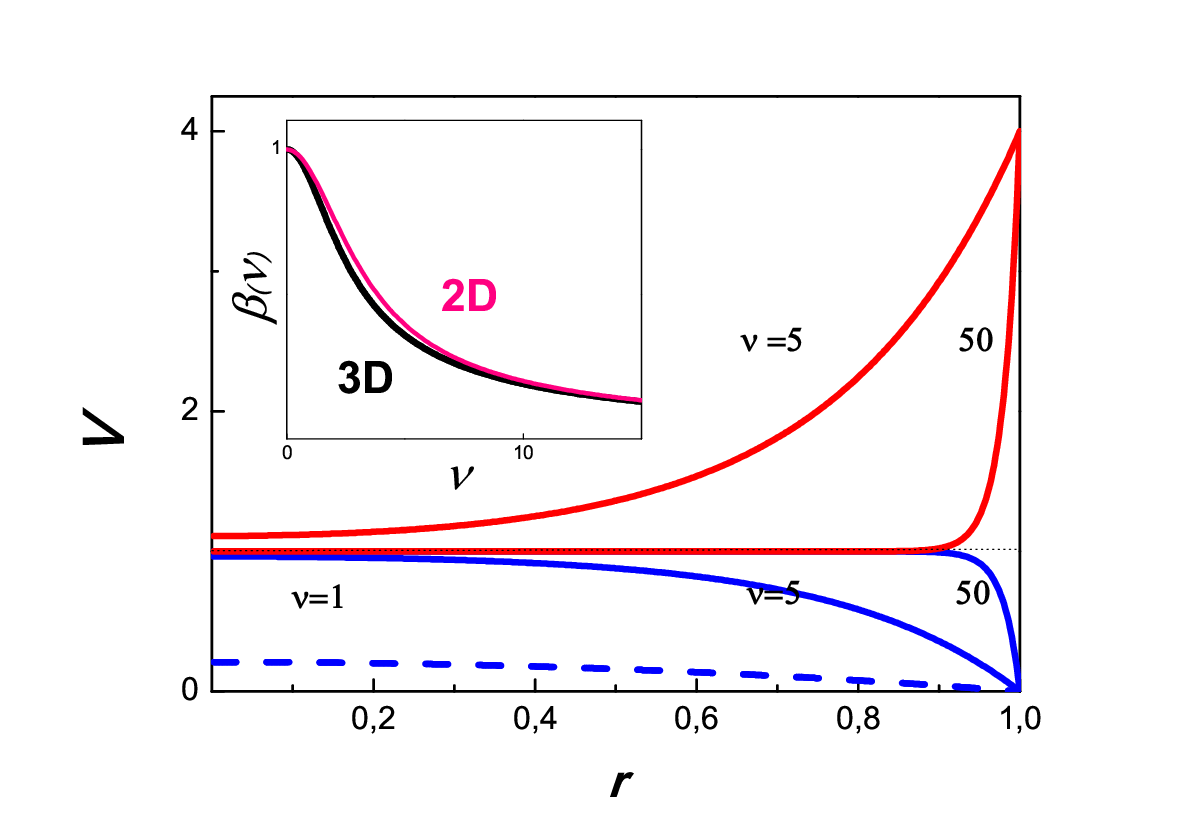} \caption[]{\label{Fig3}  Flux velocity distribution $v(r)$
specified by Eq.(\ref{Solution_Bessel2}) at fixed applied longitudinal electric field and viscosity parameter $\nu=5;50$ for
wall adhesion boundary condition $v_{0}=0$(blue) and diamagnetic boundary condition $v_{0}=4$(red).
The Poiseuille flow for $v_{0}=0;\nu=1$ is shown by the dashed line. Dotted line represents the uniform flow $v=v_{0}=1$. Inset: universal 
dependence $\beta(\mu)$ for present 3D wire case. Pink curve $\beta_{2D}(\mu)$ corresponds to 2D slab case discussed in text.}
\end{center}
\end{figure}

Our major interest concerns a strong diamagnetism case when $\kappa\geq 1$. In Fig.\ref{Fig3} we plot the radial distribution of longitudinal velocity at fixed boundary velocity $v_{0}=4$ and different strengths of the carrier viscosity. As expected, the diamagnetic current within a narrow layer $\delta$ initiates a current flow within in much wider stripe $\lambda \gg \delta$ close to sample inner wall. The flux velocity approaches a conventional ohmic drift velocity in a sample bulk.

Using Eq.(\ref{Solution_Bessel2}) one may find the average current density $j=\frac{2ne}{R_{0}^{2}}\int\limits^{R_{0}}_{0}V_{z}(R) R dR$:
\begin{equation}
j=j_{z}\left [ 1+(v_{0}-1)\beta(\nu)\right ],
\label{Average_Curent}\\
\end{equation}
where $\beta(\nu)=\frac{2J_{1}(\nu)}{\nu J_{0}(\nu)}$ is the universal function(see Fig.\ref{Fig3},inset) of the viscosity strength,
$J_{1}$ is first-order modified Bessel function of the first kind. The function $0<\beta(\nu)\leq 1$ decreases smoothly
as $\sim 1-\nu^{2}/8 $ for high-viscous case $\nu \ll 1$ and, then follows the asymptote $\sim 2/\nu$ for low viscosities $\nu \gg 1$.

Remarkably, the all previous arguments can be generalize for two-dimensional slab whose thickness is much less compared to 
other sample sizes. Actually, 2D slab thickness plays the role of a wire radius upon straightforward replacement in present notations.   
We find that for 2D slab geometry the universal function $\beta(\nu)$ can be replaced by $\beta_{2D}(\mu)=2\tanh(\nu)/\nu$ shown 
by pink line in Fig.\ref{Fig3},inset. Both dependencies are close one to each other, therefore we expect a similar effects which 
will be discussed hereafter. The detailed analysis of 2D slab case will be available elsewhere.

From Eq.(\ref{boundary_condition}) the self-consistent solution of Eq.(\ref{Average_Curent}) yields
\begin{equation}
j=j_{z}\frac{1-\beta}{1-\kappa \beta}.
\label{SC_current}\\
\end{equation}

\begin{figure}[tbp]
\begin{center}\leavevmode
\includegraphics[width=0.9\linewidth]{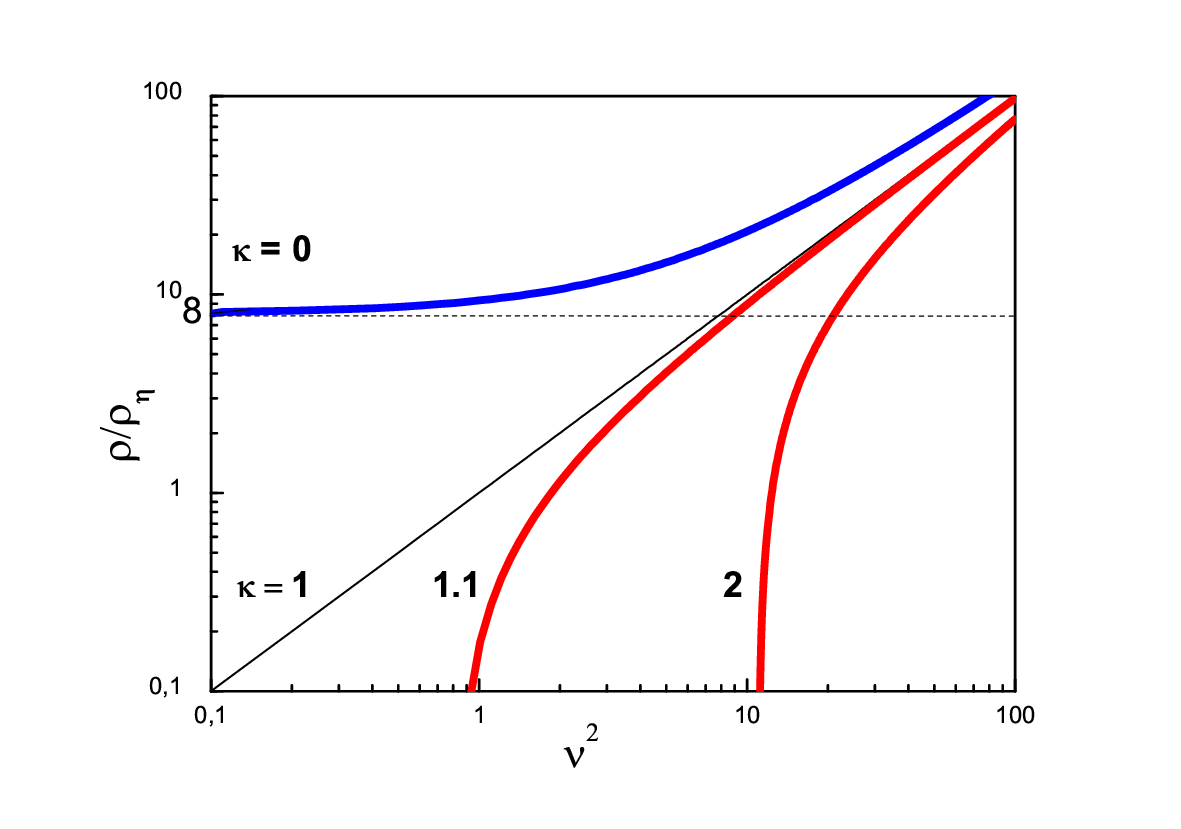} \caption[]{\label{Fig4} Dimensionless resistivity $\rho/\rho_{\eta}$ followed from
Eq.(\ref{SC_resistivity}) vs dimensionless disorder strength $\nu^{2}=\frac{R_{0}^{2}}{\eta \tau}$ for: zero
diamagnetic current $\kappa=0$; uniform current state $\kappa=1$; strong diamagnetism $\kappa>1$. Dashed line
represents the viscous resistivity $\rho=8\rho_{\eta}$ at $\kappa=0$ and $1/\tau \rightarrow 0$.}
\end{center}
\end{figure}

Eq.(\ref{SC_current}) defines the average current density at fixed longitudinal electric field $E_{z}$.
Consequently, one may define the "effective resistivity" $\rho=E_{z}/j$ as it follows
\begin{equation}
\rho=\rho_{D}\frac{1-\kappa \beta}{1-\beta},
\label{SC_resistivity}\\
\end{equation}
where $\rho_{D}=\frac{m}{ne^{2}\tau}$ is the conventional Drude resistivity.

Eq.(\ref{SC_resistivity}) represents the central result of the paper. The galvanic measurements give the "effective
resistivity" which depends on the inner wall boundary condition, sample size and, moreover, differs from expected Drude value.
At first, for $\kappa=1$ one recover the uniform current flow without viscous effects, hence $\rho=\rho_{D}$.
Secondly, the wall adhesion condition $\kappa=0$ provides the "effective
resistivity" as $\rho=\rho_{D}/(1-\beta)$ already reported in Ref.\cite{Gurzhi63}. For low-viscose case $\nu \gg 1$
the "effective resistivity" is still described by Drude formulae $\rho \sim \rho_{D}$. In the opposite high-viscosity
and(or) low dissipation limit $\nu \ll 1$ the Poiseille type of a current flow is realized. The "effective resistivity" at $\nu \ll 1$
is given by the asymptote $\rho=8\rho_{\eta}$, where $\rho_{\eta}=\frac{m}{ne^{2}}\frac{\eta}{R_{0}^{2}}$ is
so-called "viscous" resistivity\cite{Gurzhi63} which depends on the sample size.  Note, the ratio $R_{0}^{2}/\eta$ plays the role
of the momentum relaxation time similar to that discussed\cite{Alekseev16,Shi14} for 2D electron gas. The transition from Drude to "viscous"
resistivity case occurs at $\nu\sim 1$.

In Fig.\ref{Fig4} we plot the reduced resistivity $\rho/\rho_{\eta}$ vs
disorder $\nu^{2}\sim 1/\tau$ for fixed viscosity strength $\eta$ and different valued of diamagnetic parameter $\kappa$.
Note that at high disorder and(or) low viscosity $\nu \gg 1$ the resistivity in Fig.\ref{Fig4} starts to follow conventional
Drude dependence. The most intriguing feature of Eq.(\ref{SC_resistivity}) concerns the effective resistivity which may vanish at
\begin{equation}
\kappa \cdot \beta(\nu)=1.
\label{Critical_condition}\\
\end{equation}
Eq.(\ref{Critical_condition}) gives the critical condition for so-called "zero-resistance state" (ZRS) seems to be
observed in Ref.\cite{Kamerlingh-Onnes1911}. Recall that for arbitrary argument $\beta(\nu) \leq 1$. Thus, the solution
of transcendental Eq.(\ref{Critical_condition}) is possible when $\kappa>1$. The condition $\kappa>1$ can be re-written
as $R_{0} \geq R_{m}$, where we introduce the minimal wire radius
\begin{equation}
R_{m}=\frac{\delta}{2\pi |\chi|},
\label{Minimal_wire_radius}\\
\end{equation}
when the ZRS can be realized. We further demonstrate that ZRS criteria $\kappa=R_{0}/R_{m}>1$ could be even stronger
regarding real systems.

We emphasize that the zero resistance state may appear for even finite momentum relaxation time. At a first glance,
this result looks like mysterious. Nevertheless, the experimental data\cite{Meissner32} provide a strong evidence of
the disorder remains indeed finite within zero-resistance state. We argue that the physics of zero-resistance state is
rather transparent. The non-dissipative diamagnetic current is pinched within a narrow inner layer $\lambda \sim R_{m}/2$
of a wire and, then shunts the dissipative current in the sample bulk. The total current in a
wire becomes purely diamagnetic when Eq.(\ref{Critical_condition}) is fulfilled.

\subsection{Size-dependent transition to zero-resistance state}
\label{Size-dependent zero-resistance state transition}
\begin{figure}[tbp]
\begin{center}\leavevmode
\includegraphics[width=1.0\linewidth]{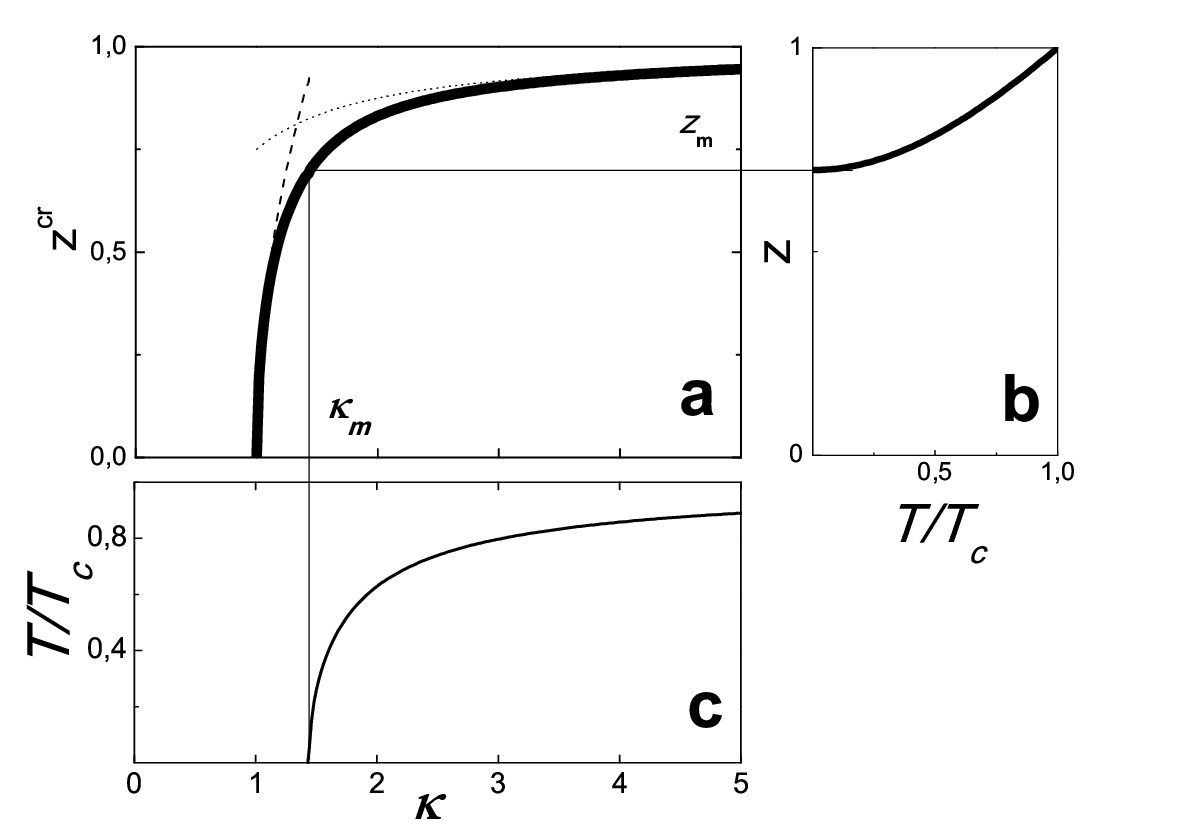} \caption[]{\label{Fig5} a) The critical diagram $z^{cr}(\kappa)$ of zero-resistance
state followed from Eq.(\ref{transcendental_equation}). The asymptotes for small $\kappa-1\ll 1$ and bulky sample $\kappa \gg 1$ are shown by dashed and dotted line respectively. The area below the critical curve $z^{cr}(\kappa)$ corresponds to zero-resistance state. b) The dependence $z(\Theta)$ specified by Eq.(\ref{z(T)plus}). c) The resulting temperature threshold dependence $\Theta(\kappa)$ specified by Eq.(\ref{Threshold_temperature}) for fixed $z_{m}=0.7$.}
\end{center}
\end{figure}

We now examine in greater details the critical condition given by Eq.(\ref{Critical_condition}). One can find, in principle,
the critical dependence in a following form $\nu^{cr}(\kappa)$. The latter is, however, non-informative since both
variables $\kappa,\nu$ depend on the sample size. To avoid this problem, let us introduce a size-free parameter
$z=\frac{\nu}{2\kappa}=\frac{R_{m}}{2\lambda}$. The modified Eq.(\ref{Critical_condition}) yields the transcendental equation
\begin{equation}
z=\frac{J_{1}(2\kappa z)}{J_{0}(2\kappa z)}.
\label{transcendental_equation}\\
\end{equation}
which gives a desired critical diagram in a form $z^{cr}(\kappa)$. The latter is shown in Fig.\ref{Fig5},a. Again, the area below the critical curve
corresponds to zero resistance state. For sample size closed to its minimal value $R_{m}$, i.e. when $\kappa-1\ll 1$, the critical curve follows the asymptote $z^{cr}=\sqrt{2(\kappa-1)}$ depicted by the dashed line in Fig.\ref{Fig5},a. Then, the critical curve saturates asymptotically as $z^{cr}(\kappa)=1-\frac{1}{4\kappa}$ for bulky sample, i.e. when $\kappa\gg 1$.

Up to this moment we assumed the momentum relaxation time and the carrier viscosity to be temperature independent.
One can make an attempt to find ZRS threshold in terms of temperature since $z\sim 1/\lambda=1/\sqrt{\eta\tau}$. Remind that for actual low-T case
the transport is mostly governed by scattering on static defects, hence one may consider the T-independent momentum relaxation time
$\tau \neq \tau(T)$. In contrast, the e-e scattering time is known\cite{Pomeranchuk50,Abrikosov59,Baym67} to be a strong function
of temperature. Thus, we assign
\begin{equation}
\frac{1}{\tau_{ee}(T)}=\frac{\xi^{2}}{\tau^{1}_{ee}}+\frac{1}{\tau^{0}_{ee}},
\label{ee_scattering_time}\\
\end{equation}
where $\xi=T/T_{F}$ is the degeneracy parameter, $T_{F}=\varepsilon_{F}/k$ and $\varepsilon_{F}$ are the Fermi temperature and energy
respectively. Then, $\tau^{0}_{ee}$ is the residual value
of e-e scattering time at $T \rightarrow 0$, $\tau^{1}_{ee}$ is a dimensional value known to be of the
order of $\hbar/\varepsilon_{F}$ within Fermi liquid theory\cite{Pines96}. In general, both values $\tau^{0,1}_{ee}$ are unknown, thus
stay to be extracted from experimental data.

With the help of Eq.(\ref{ee_scattering_time}) the parameter $z=\frac{R_{m}}{2\lambda}$ becomes temperature dependent, namely
\begin{equation}
z(\xi)= z_{m} \sqrt{1+\gamma \xi^{2}},
\label{z(T)}\\
\end{equation}
where $\gamma=\tau^{0}_{ee}/\tau^{1}_{ee}$ is the dimensionless ratio, then $z(0)=z_{m}=\frac{\sqrt{5}R_{m}}{2v_{F}\sqrt{\tau^{0}_{ee}\tau}}$
is the zero temperature value. Recall that $z^{cr}(\kappa)\leq 1$, hence a condition $z_{m} \leq 1$ must be satisfied. The latter
gives the condition
\begin{equation}
\mu\geq\mu_{\text{min}},
\label{condition_Zm}\\
\end{equation}
for carrier mobility, where $\mu_{\text{min}}= \frac{5}{8}\frac{eR_{m}^{2}}{\varepsilon_{F}\tau^{0}_{ee}}$ plays the role of the minimal
mobility for which the ZRS is possible. Further, we will use a trivial relationship $z_{m}=\sqrt{\mu_{\text{min}}/\mu}$ as well.

If $z_{m}<1$, the only upper part $z^{cr}(\kappa)>z_{m}$ of the
threshold diagram in Fig.\ref{Fig5},a remains useful. Then, the equality $z_{m}=z^{cr}(\kappa_{m})$ denotes a certain value
of minimal sample size parameter $\kappa_{m}$, which corresponds to ZRS threshold at $T=0$. Evidence shows that at finite
temperature the zero resistance state can be realized for samples whose sizes satisfy the condition $\kappa\geq\kappa_{m}$.
The latter gives the strict criteria
\begin{equation}
R_{0}\geq R_{m}\cdot \kappa_{m}
\label{minimal_sample_size}\\
\end{equation}
for minimal sample radius instead of that $R_{0}\geq R_{m}$ discussed earlier.

We now attempt to find out the threshold temperature for massive sample( i.e. when $\kappa \gg 1$ ) known to be a
universal quantity\cite{Kamerlingh-Onnes1911} for certain material. Helpfully, it can be done within our model. Indeed, with the help
of Eq.(\ref{z(T)}) and condition $z(\xi)=1$ valid for massive sample one obtains the subsequent threshold temperature $T_{c}$:
\begin{equation}
T_{c}= T_{F} \left[ (z^{-2}_{m}-1)/\gamma \right]^{1/2}.
\label{T_c}\\
\end{equation}
Hereafter, we will label the all quantities related to ZRS threshold in bulky sample by index "c".
Remarkably, one may re-write Eq.(\ref{T_c}) in terms of resistivity $\rho_{c}=\rho|_{T_{c}}$
associated with the critical temperature $T_{c}$:
\begin{equation}
\frac{\rho_{c}}{\rho_{\text{max}}}= \frac{1}{1+\gamma (T_{c}/T_{F})^{2}},
\label{T_c_resistivity}\\
\end{equation}
where we use a notation $\rho_{\text{max}}=(ne\mu_{\text{min}})^{-1}$. Let us assume a pure massive sample exhibited a certain
ZRS threshold point at $\rho_{c},T_{c}$. The momentum relaxation time may depend on static
disorder. Eq.(\ref{T_c_resistivity}) provides an evidence of ZRS threshold temperature decay caused by
disorder enhancement. These predictions is qualitatively confirmed by experimental observations\cite{Lynton57}.

It is useful to introduce the reduced temperature $\Theta=T/T_{c}$. Consequently, Eq.(\ref{z(T)}) can be modified as it follows
\begin{equation}
z(\Theta)= \sqrt{z_{m}^{2}+(1-z_{m}^{2})\Theta^{2}},
\label{z(T)plus}\\
\end{equation}
and, then plotted in Fig.\ref{Fig5},b. Combining the dependencies $z(\Theta)$ and $z^{cr}(\kappa)$ specified by Eq.(\ref{z(T)plus})
and Eq.(\ref{transcendental_equation}) respectively one obtains threshold temperature as a function of the sample size $\Theta(\kappa)$:
\begin{equation}
\Theta(\kappa) = \left[ \frac{z^{cr}(\kappa)^{2}-z^{2}_{m}}{1-z^{2}_{m}} \right]^{1/2}.
\label{Threshold_temperature}\\
\end{equation}
An example is shown in Fig.\ref{Fig5},c. Experimentally, threshold temperature diminution was
observed\cite{Meissner33,Meissner35,Burton34} for small sized samples.

\begin{figure}[tbp]
\begin{center}\leavevmode
\includegraphics[width=1.2\linewidth]{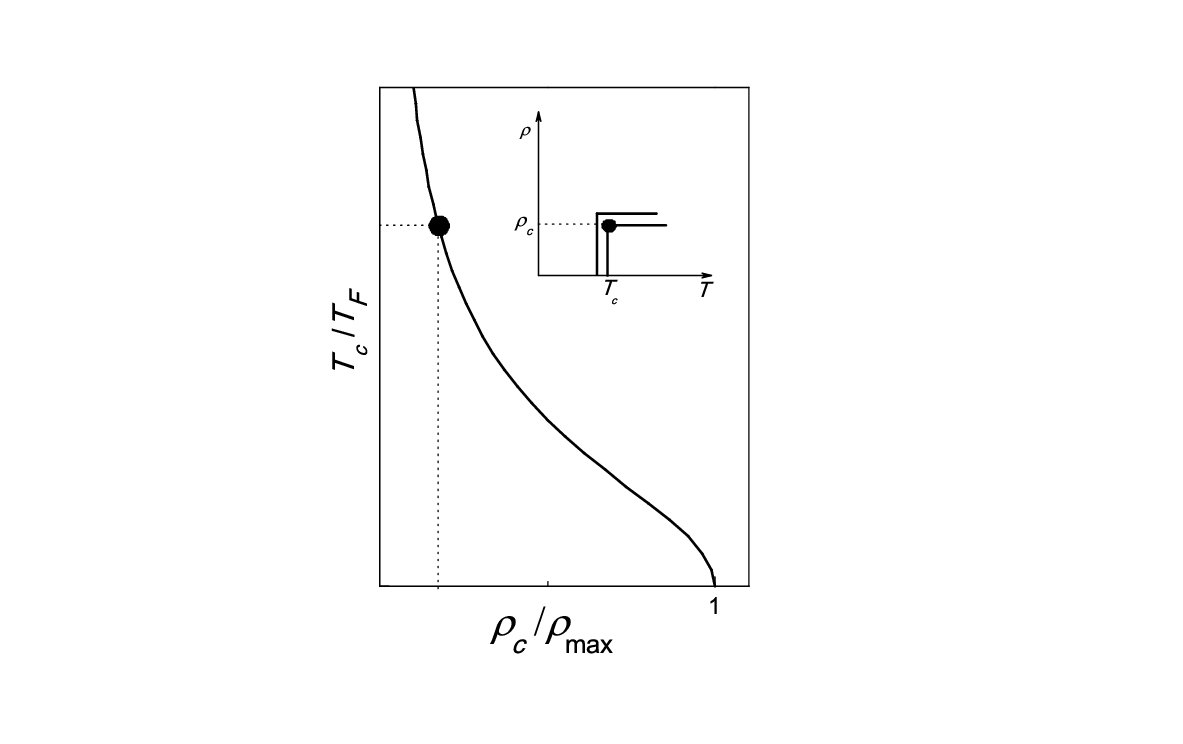} \caption[]{\label{Fig5a} a) Dependence $\rho_{c}(T_{c})$
specified by Eq.(\ref{T_c_resistivity}). Inset: sketch view of experimental resistivity curves $\rho(T)$
used to determine $\rho_{c}(T_{c})$.}
\end{center}
\end{figure}

\begin{figure}[tbp]
\begin{center}\leavevmode
\includegraphics[width=1.0\linewidth]{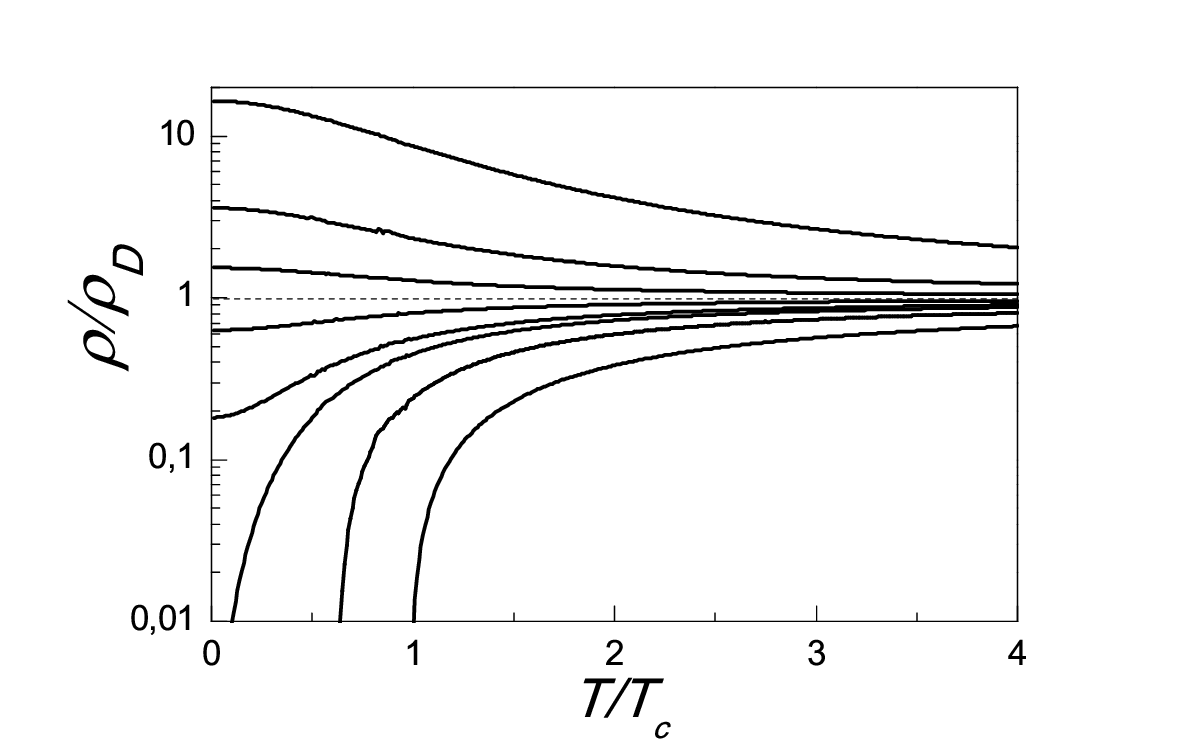} \caption[]{\label{Fig5aa} a) Resistivity set $\rho(\Theta)$ for sample-size driven ZRS-to-normal state transition for $z_{m}=0.7$ and $k=R_{0}/R_{m}=30;2;k_{m}=1.438;1.3;1.1;1;0.9;0.7;0.4$(from bottom to top).}
\end{center}
\end{figure}

Finally, we will explore our model in order to demonstrate a possibility of sample-size driven ZRS to normal
metal transition. Let us consider a wire for which the zero-resistance state can be realized. For example, we assign $z_{m}=0.7<1$.
With the help of critical diagram shown in Fig.\ref{Fig5} we find the minimal value of diamagnetic parameter $\kappa_{m}=1.44$.
Using Eq.(\ref{z(T)plus}) and, then substituting $\nu=2\kappa z(\Theta)$ into Eq.(\ref{SC_resistivity}) one obtains
T-dependent resistivity $\rho(\Theta)$ for fixed values of sample radius $\kappa=R_{0}/R_{m}$ . The result is shown in Fig.\ref{Fig5aa}.
Evidence shows that the change from apparent "metallic" $\frac{d\rho}{dT}>0$ to "insulating" $\frac{d\rho}{dT}<0$ behavior
occurs when $\kappa=1$, i.e. for uniform current flow regime. We claim that the key parameter $R_{m}$ can
be amenable for experimental test regarding divergent $\rho(T)$-data analysis.

\subsubsection{Proximity Effect}
\begin{figure}[tbp]
\begin{center}\leavevmode
\includegraphics[width=1.0\linewidth]{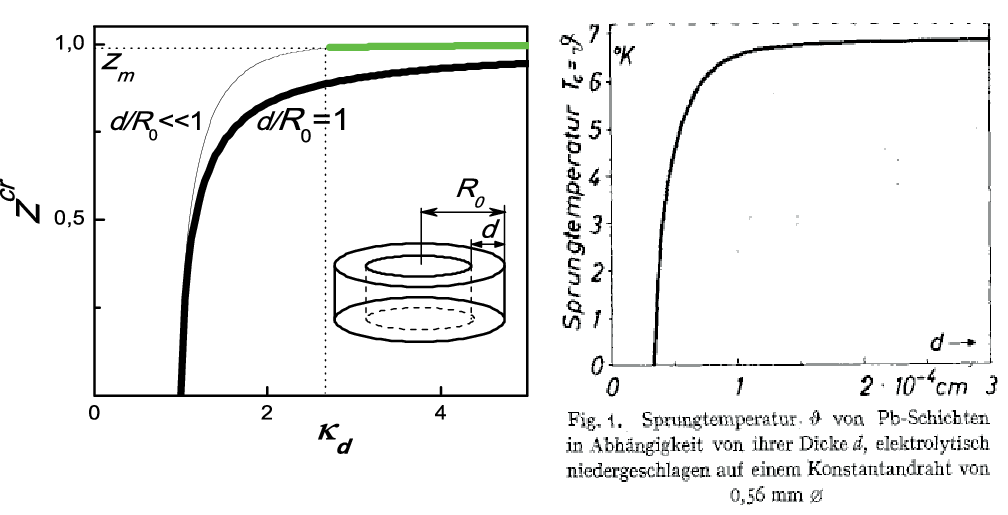} \caption[]{\label{Fig5b} a) The critical diagram $z^{cr}(\kappa_{d})$ followed from Eq.(\ref{transcendental_equation_ring}) for solid wire $d/R_{0}=1$ and thin coating layer $d/R_{0} \ll 1$. Green curve depicts the
experimental data in the panel b. Insert: coaxial sample geometry. b) Critical temperature vs Pb layer thickness  for
fixed inner core radius $0.28\text{mm}$ under Ref.\cite{Hilsch62}.}
\end{center}
\end{figure}

In general, the experimental observation of the proposed size-dependent threshold transition could be rather difficult since it requires a typical
wire radius of the order of hundred angstroms. To avoid this problem the authors of Refs.\cite{Burton34,Misener35,Hilsch62} suggested a coaxial construction of normal metal kernel( see Fig.\ref{Fig5b}a, inset ) of a fixed radius coated by Pb-superconductor layer of angstrom scale width $d\ll R_{0}$. The layer width $d$ can be varied. In Fig.\ref{Fig5b}b we reproduce the  critical temperature $T(d)$ data\cite{Hilsch62} for coaxial sample. The superconducting state occurs when the coating width exceeds a certain minimal value $d_{0}$. We now demonstrate that this result can be easily obtained within our model. Indeed, we reproduce the all previous calculations for coaxial geometry and, then obtain the following equation
\begin{equation}
z=\frac{Y_{0}(\nu\Delta)\left [ J_{1}(\nu)-\Delta J_{1}(\nu\Delta)\right ]+J_{0}(\nu\Delta)\left [Y_{1}(\nu)-\Delta Y_{1}(\nu\Delta)\right ]}{Y_{0}(\nu\Delta)J_{0}(\nu)- Y_{0}(\nu)J_{1}(\nu\Delta)},
\label{transcendental_equation_ring}\\
\end{equation}
where $\Delta=1-d/R_{0}$ is the dimensionless ratio of the inner core radius to $R_{0}$, then $Y_{1}$ is the first-order modified Bessel function of the second
kind. Surprisingly, Eq.(\ref{transcendental_equation_ring}) does not contain any component caused by dissipative current inside the metallic core.
We find that even the kernel is empty the Eq.(\ref{transcendental_equation_ring}) remains unchanged. Indeed, our previous findings specified by Eq.(\ref{transcendental_equation}) deals with a purely non-dissipative diamagnetic current flow in a narrow layer nearby the inner
wall of a wire. The all dissipative currents can be disregarded in this respect. The same reasoning is valid for present case of coaxial sample
with(without) the normal metal inside the core. Therefore, Eq.(\ref{transcendental_equation_ring}) is universal.

For fixed value of the core radius $\Delta \cdot R_{0}$ the solution of Eq.(\ref{transcendental_equation_ring})
provides a set of critical diagram curves $z^{cr}(\kappa_{d})$. Here, we make use of the dimensionless Pb-layer width
$\kappa_{d}=d/R_{m}$ similar to variable $\kappa$ used above for simple wire case. The result is shown in Fig.\ref{Fig5b}a.
The zero-resistance state is possible when $\kappa_{d}>1$. For coreless wire $\kappa_{d}\rightarrow \kappa$ we
readily reproduce the critical diagram shown previously in Fig.\ref{Fig5}. In the opposite case of massive core
coated by a thin layer, i.e. when $d/R_{0} \ll 1$, the critical diagram is upshifted( see Fig.\ref{Fig5b}a).

We now compare our model finding with experiment\cite{Hilsch62}.
Note that the present critical diagram $z^{cr}(\kappa_{d})$ can be used to deduce the critical temperature dependence
$\Theta(\kappa_{d})$ similar to that depicted in Fig.\ref{Fig5}b,c. On the contrary, one may use the experimental
dependence $\Theta(d)$ and, then impose it to a appropriate curve on the threshold diagram set $z^{cr}(\kappa_{d})$.

Keeping $z_{m}$ and $R_{m}$ as a fitting parameters, we plot in Fig.\ref{Fig5b}a the dimensionless replica of the experimental data
shown in Fig.\ref{Fig5b}b. For actual core radius $0.28$mm and minimal coating width $d_{0}=3500$A\cite{Hilsch62} our
best fit gives $z_{m}=0.99$ and $d_{0}/R_{m}=2.71$. Therefore, we obtain $R_{m}=1290$A.

\subsection{Magnetic field screening}
\label{Magnetic field screening}
\begin{figure}[tbp]
\begin{center}\leavevmode
\includegraphics[width=0.9\linewidth]{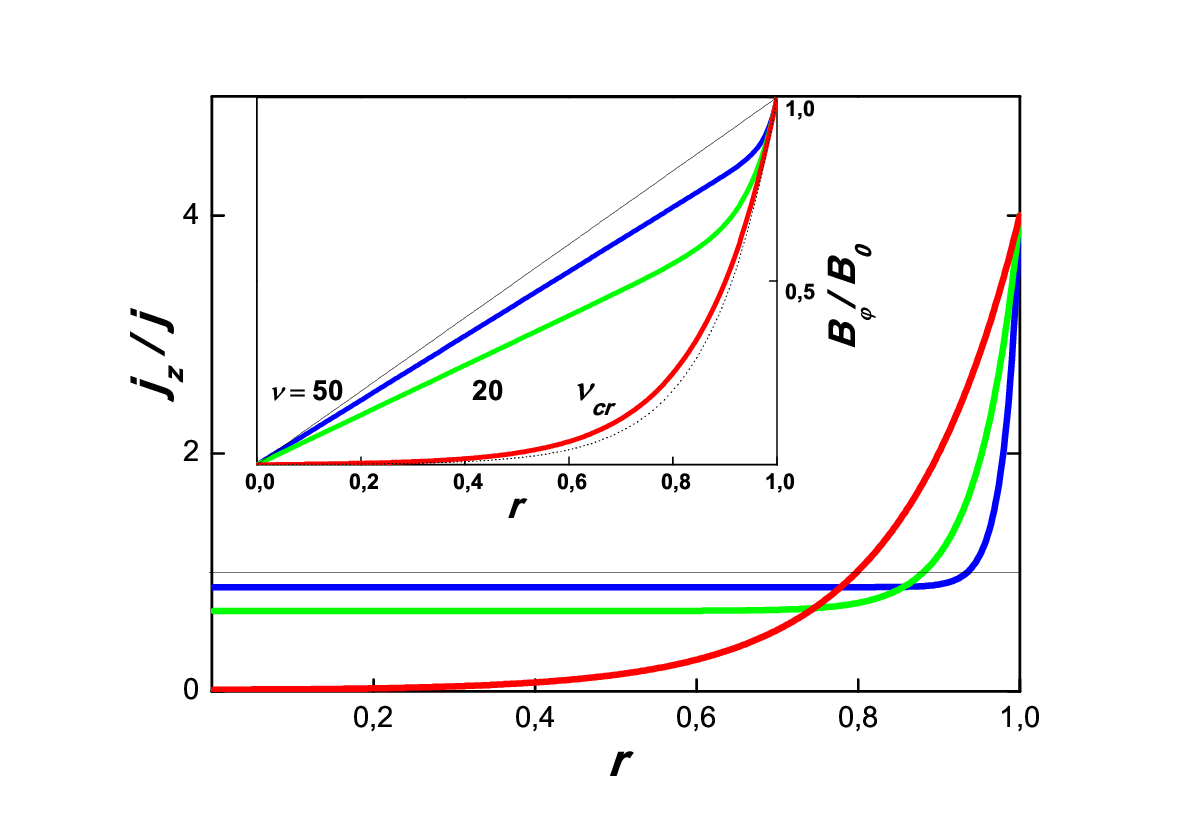} \caption[]{\label{Fig6} Distribution of the dimensionless current density $j_{z}/j$ and azimuthal magnetic field $B_{\varphi}/B_{0}$ (insert) specified by Eq.(\ref{J_B real}) for finite size sample at $\kappa=4$(corresponds to $z_{cr}=0.93$ and
$\nu_{cr}=2\kappa z_{cr}=7.44$) and viscosity parameter $\nu=\nu_{cr};20;50$. Thin lines depict the uniform current density case when $\kappa=1$. Dotted line(insert) corresponds to magnetic field screening asymptote described in text.}
\end{center}
\end{figure}

We now demonstrate that the magnetic field can be pushed out from the sample bulk as stronger as the system becomes closer to
zero resistance state threshold. Remind that the flux velocity distribution specified by Eq.(\ref{Solution_Bessel2}) was found under
assumption of a fixed electric field $E_{z}$. Using Eq.(\ref{SC_current}) the later can be represented in terms of total current $I$.
As a result, both the radial distribution of the current density $j_{z}(r)$ and the azimuthal magnetic field $B_{\varphi}(r)$ specified by Eqs.(\ref{Solution_Bessel2}),(\ref{boundary_condition}) and Eq.(\ref{Azimuthal_Field}) respectively yield
\begin{eqnarray}
j_{z}(r)=j \left [ \frac{1-\kappa \beta}{1-\beta}+\frac{\kappa-1}{1-\beta} \cdot \frac{J_{0}(r\nu)}{J_{0}(\nu)} \right ],
\eqnum{1} \label{J_B real}\\
B_{\varphi}(r)=B_{0}\left [\frac{1-\kappa \beta}{1-\beta}r+\beta \cdot \frac{\kappa-1}{1-\beta}\cdot \frac{J_{1}(r\nu)}{J_{1}(\nu)} \right ].
\nonumber
\end{eqnarray}
Remind that $j=\frac{I}{\pi R_{0}^{2}}$ is the average current density.
As expected, for uniform flow $\kappa=1$ one obtains $j_{z}(r)=j$, $B_{\varphi}=B_{0}r$. Then, Eq.(\ref{J_B real}) gives the correct values of
the current density $j_{z}(1)=j\kappa$ and the magnetic field $B_{\varphi}(1)=B_{0}$ at the inner surface of the wire.
We plot the dependencies given by Eq.(\ref{J_B real}) in Fig.\ref{Fig6}. At fixed diamagnetic parameter $\kappa>1$ the growth of
the fluid viscosity leads to progressive shift of the current towards the inner wall of the wire. Simultaneously, the magnetic field
is pushed out from the sample bulk.

Remind that the typical length scale of viscose flow yields $\lambda=\sqrt{\eta\tau}=\frac{R_{m}}{2z}$. For bulky sample at ZRS threshold
$z=1$ one obtains the screening length $\lambda= l_{B}= R_{m}/2$. As an example, we plot in Fig.\ref{Fig6}, inset the magnetic field
screening asymptote $B_{\varphi}=B_{0}\exp\left ({\frac{R-R_{0}}{l_{B}}} \right )$.

Our final remark concerns the presence of the radial electric field. With the help of Eq.(\ref{YX_velocity_Stokes}) and Eq.(\ref{J_B real}) we obtain $E_{R}=\frac{j_{z}(r)B_{\varphi}(r)}{nec}$. Following our previous arguments the longitudinal current in a sample bulk can be viewed as a carriers drift in a crossed $E_{R}\perp B_{\varphi}$ fields.

\subsection{Magnetic field phase diagram of zero-resistance state}
\label{Magnetic field phase diagram of zero-resistance state}
Remind that the all previous discussion concerned the zero-current limit of the transport measurements. The critical
diagram of zero-resistance state was found for arbitrary sample size.
In reality, a finite applied current and, hence accompanied current-induced azimuthal magnetic field are known\cite{Meissner33,Meissner33+,Silsbee17}
to influence threshold temperature of zero-resistance state. In order to account for current driven effects one must solve Eq.(\ref{Vz_Stokes_Modified})
modified with respect to magnetic field dependent longitudinal viscosity specified by Eq.(\ref{viscosity_components}). The resulting equation
is rather difficult to be analytically resolved. However, one may qualitatively catch the underlying physics. For simplicity, we restrict
ourself to massive sample case.

We re-write the threshold criteria given by Eq.(\ref{z(T)}) with carrier viscosity $\eta_{xx}$ included.
\begin{equation}
z(\xi,\Omega_{c})=z_{m} \sqrt{1+\gamma \xi^{2}}\sqrt{1+\frac{4\Omega_{c}^{2}(\tau_{ee}^{0})^{2}}{(1+\gamma \xi^{2})^{2}}}.
\label{z(T,B)}\\
\end{equation}
Note that within the above approach we neglect, in fact, the magnetic field radial distribution in a sample bulk.

To proceed, we write down a criteria $z(\xi,\Omega_{c})=1$ for massive specimen and obtain the threshold diagram
in terms of the magnetic field vs temperature:
\begin{equation}
\frac{\Omega_{c}}{\Omega_{c}(0)}=\sqrt{(1-\Theta^{2})(1+(z_{m}^{-2}-1)\Theta^{2})}
\label{Diagram_B_T}\\
\end{equation}
where
\begin{equation}
\Omega_{c}(0)=\frac{\sqrt{z^{-2}_{m}-1}}{2\tau_{ee}^{0}}=\frac{1}{2\sqrt{\tau_{ee}^{0}\tau_{ee}^{1}}}\frac{T_{c}}{T_{F}}
\label{Critical_magnetic_field}\\
\end{equation}
is the cyclotron frequency and $B_{c}=mc\Omega_{c}(0)/e$ is the critical magnetic field respectively at $T\rightarrow 0$.
Eq.(\ref{Critical_magnetic_field}) confirms the proportionality $B_{c}\sim T_{c}$ observed in experiment.

\begin{figure}[tbp]
\begin{center}\leavevmode
\includegraphics[width=0.9\linewidth]{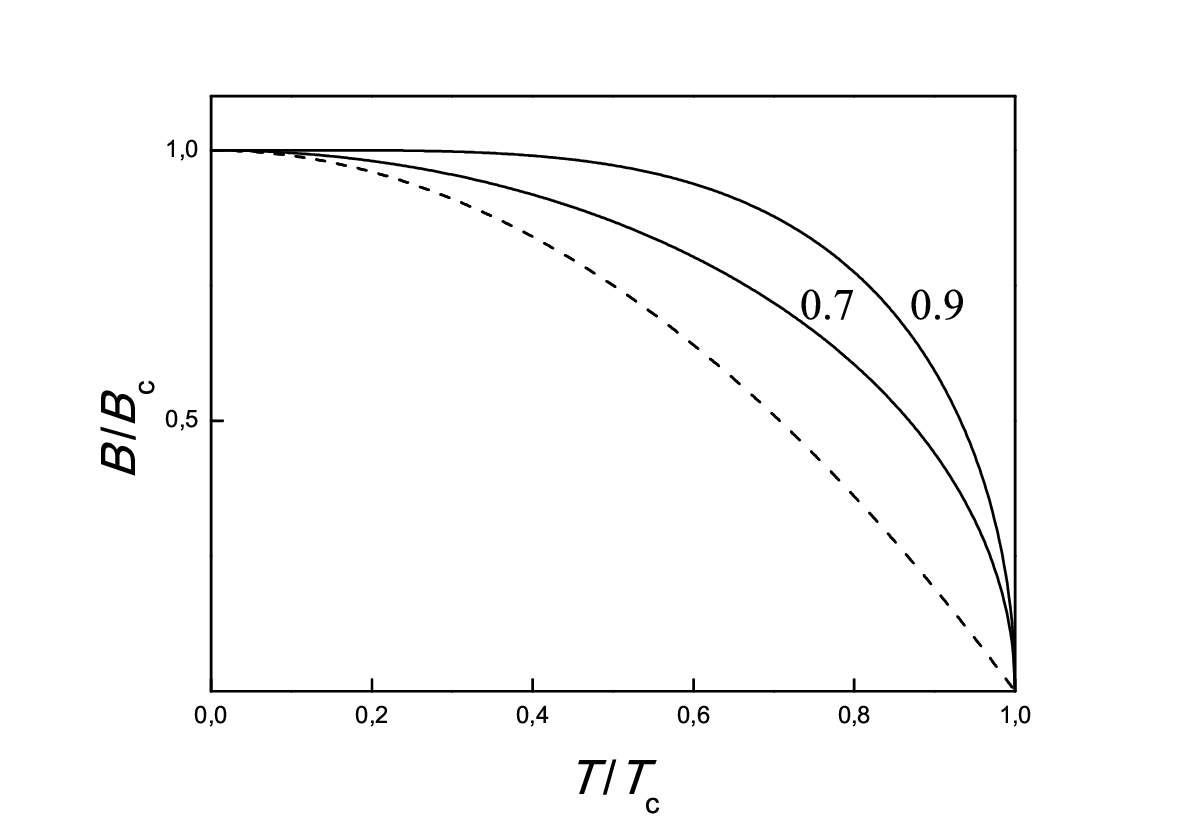} \caption[]{\label{Fig7} The critical magnetic field vs critical temperature specified by Eq.(\ref{Diagram_B_T}) for massive sample at $z_{m}=0.7; 0.9$. The dashed curve corresponds to empirical dependence $B/B_{c}=1-\Theta^{2}$.}
\end{center}
\end{figure}

As an example, the magnetic field driven threshold diagram specified by Eq.(\ref{Diagram_B_T}) is plotted in Fig.\ref{Fig7}. The critical curve has a quadrant shape which is close to empiric dependence $B/B_{c}=1-\Theta^{2}$ often used in practice.

\subsection{Estimations}
\label{Resistance modeling}
For the sake of certainty, we further examine the massive lead whose typical low-T resistivity data
data\cite{Clusius32} is represented in Fig.\ref{Fig8}. Let us use textbook values\cite{Pool2007} of Fermi energy $\varepsilon_{F}=9.47$eV, velocity $v_{F}=1.83\times 10^{8} \text{cm/s}$ and carrier density $n=1.3\times 10^{23} \text{cm}^{-3}$ of a massive lead. At $T=273$K the typical resistivity is $\rho=1.9 \times 10^{-7}\Omega \text{m}$\cite{Pool2007}. Hence, the low-T data\cite{Clusius32} denote the resistivity $\rho_{c}=3.8 \times 10^{-10}\Omega \text{m}$ at ZRS threshold $T_{c}=7.2$K. The respective carrier mobility $\mu=1/(ne\rho_{c})=1240\text{cm}^{2}/\text{Vs}$ allows one to estimate the momentum relaxation time $\tau=7.1 \cdot 10^{-13}\text{s}$ and the transport length $l=v_{F}\tau=1.3\mu\text{m}$. Using the above extracted value $z_{m}=0.99$ we calculate the minimal mobility as $\mu_{\text{min}}=\mu z^{2}_{m}=1215\text{cm}^{2}/\text{Vs}$. Our previous finding $R_{m}=1290$A
gives the residual e-e scattering time $\tau^{0}_{ee}=\frac{5eR^{2}_{m}}{8\epsilon_{F}\mu_{\text{min}}}=0.9\cdot 10^{-14}\text{s}$. Then, we are able to
calculate a ratio $\gamma=(z^{-2}_{m}-1)(T_{F}/T_{c})^{2}=4.7\cdot 10^{6}$ and, finally, deduce $\tau^{1}_{ee}=1.9\cdot 10^{-21}\text{s}$ imbedded into Eq.(\ref{ee_scattering_time}). Note that for actual temperatures $\sim T_{c}$ the total e-e scattering time specified by Eq.(\ref{ee_scattering_time})
is mostly determined by residual component, therefore $\tau_{ee}\sim \tau^{0}_{ee}$. The estimation of e-e scattering length $l_{ee}\sim v_{F}\tau^{0}_{ee}=0.02\mu\text{m}$ justifies the applicability of hydrodynamic approach since $l_{ee} \ll l$. It is instructive to find the magnetic filed penetration length at ZRS threshold as $l_{B}=R_{m}/2=640$A being of the order of magnitude of that $390$A known in literature\cite{Lynton1971}. Using Eq.(\ref{Critical_magnetic_field}) we find also the cyclotron frequency $\Omega_{c}(0)=7.8\times 10^{12}\text{c}^{-1}$ at $T\rightarrow 0$. The respective critical magnetic field $B_{c}=4.4\cdot 10^{5}$G is, however, much higher than that $\sim 803$G observed experimentally. We attribute the above discrepancy
to approximate analytic approach used to find threshold criteria in presence of finite magnetic field.

Following Landau's theory let us estimate the diamagnetic susceptibility of free electron gas $\chi=-\frac{1}{2}\frac{n \mu_{B}^{2}}{\varepsilon_{F}}=3.8\cdot 10^{-7}$, where $\mu_{B}=\frac{e\hbar}{2mc}$ is the Bohr's magneton. The diamagnetic current is caused by movement of carriers on skipping orbits whose deviation from a sample wall $\delta=2\pi |\chi|R_{m}=4.6\cdot 10^{-13}$m is less than the average interelectronic distance $r_{s}=(4\pi n/3)^{-1/3}=1.2$A.

\subsection{Conclusions}
In conclusion, we discover the Circular Hall Effect in a wire taking into account both the diamagnetism and finite viscosity
of 3D electron liquid. We demonstrate that under certain condition the resistivity of the sample vanishes exhibiting the transition
to zero-resistance state. The to current is pinched nearby the inner rod boundary while the magnetic is pushed out of the sample bulk.
Within low current limit the threshold temperature is calculated for arbitrary carrier dissipation and the sample size. For sample size and(or)
carrier mobility which are lower than a certain minimum values the zero-resistance state cannot be realized. For massive sample the
account of finite currents makes it possible to find out the threshold diagram in terms of magnetic field vs temperature.


\begin{references}
\bibitem{Hall1879} E.H. Hall, American Journal of Mathematics, {\bf 2} , 287 (1879)
\bibitem{Anselm78} A.I. Anselm, Vvedenie v fiziku poluprovodnokov, Moskow, Nauka 616p, (1978)
\bibitem{Matzek68} M.A. Matzek and B.R. Russell, Am.J.Phys. 36, 905 (1968)
\bibitem{McDonald2010}Kirk T.McDonald, http://www.physics.princeton.edu/~mcdonald/examples/wire.pdf
\bibitem{Steinberg58} M.S. Steinberg, Phys.Rev. {\bf 109}, 1486 (1958)
\bibitem{Alekseev16}P.S. Alekseev, Phys.Rev.Lett. {\bf 117}, 166601 (2016)
\bibitem{Gurzhi63} R.N. Gurzhi, Sov.Phys.JETP, {\bf 17}, 521 (1963)
\bibitem{Sivukhin96} D.V. Sivukhin, A Course of General Physics, vol. III, Electricity,
3rd Edn., Nauka, Moskow( in Russian), (1996)
\bibitem{Vlasov05} A.A. Vlasov, Makroskopicheskaya Elektrodynamika, Moskow, Fizmatlit, 240p, (2005)
\bibitem{Shi14} Q.Shi et al, Phys.Rev.B, {\bf 89}, 201301(R) (2014)
\bibitem{Kamerlingh-Onnes1911} H. Kamerlingh Onnes, Communication from the Physical Laboratory of the
University of Leiden, 122b, 124c (1911); 133a, 133c (1913)
\bibitem{Meissner32} W. Meissner, Ann. Physik (5) {\bf 13}, 641 (1932)
\bibitem{Pomeranchuk50} I.Ia.Pomeranchuk, J.Exp.Theor.Phys. {\bf 20}, 919 (1950)
\bibitem{Abrikosov59} A.A.Abrikosov and I.M.Khalatnikov, Rep.Prog.Phys.{\bf 22}, 329 (1959)
\bibitem{Baym67} G.Baym and C.Ebner, Phys.Rev.{\bf 164}, 235 (1967)
\bibitem{Pines96} D. Pines and P. Nozieres, The Theory of Quantum Liquids, Benjamin,
New York, 1996, Vol. 1.
\bibitem{Meissner33} W.Meissner, Physics-Uspekhi {\bf 13}, 639 (1933)
\bibitem{Meissner35} W.Meissner, Phys. Z. {\bf 35}, 931 (1934)
\bibitem{Burton34} E.F.Burton, J.O.Wilhelma and A.D.Misener, Trans.Roy.Soc. Canada {\bf 28}, 65 (1934)
\bibitem{Lynton57} E.A.Lynton, B.Serin  and M.Zucker, J. Phys. Chem. Solids {\bf 3}, 165 (1957)
\bibitem{Misener35} A.D.Misener and J.O. Wilhelm, Trans.Roy.Soc. Canada {\bf 29} 5 (1935)
\bibitem{Hilsch62} P.Hilsch, Zeitschrift fur Physik {\bf 167}, 511 (1962)
\bibitem{Meissner33+} W.Meissner, R.Ochsenfeld, Naturwissenschaften {\bf  21}, 787 (1933)
\bibitem{Silsbee17}F.B.Silsbee, J.Franklin Inst. {\bf 184}, 111 (1917)
\bibitem{Clusius32}Von K.Clusius, Z.Elektrochem., {\bf 38}, 312 (1932)
\bibitem{Pool2007}Ñ.Pool et al, Superconductivity, Academic Press, Amsterdam, (2007)
\bibitem{Lynton1971}E.A.Lynton, Superconductivity, Chapman and Hall, London, (1971)
\end{references}
\end{document}